\documentclass[prl,showpacs,twocolumn]{revtex4}

\usepackage{graphicx}
\usepackage{bm}
\usepackage{pifont}
\usepackage{amsmath}

\newcommand {\etal}{{\it et al.}}

\newcommand{\COSO}{Cu$_2$OSeO$_3$}

\begin{document}

\title{Magnetoelectric nature of skyrmions in a chiral magnetic insulator Cu$_2$OSeO$_3$}

\author{S. Seki$^1$, S. Ishiwata$^1$, and Y. Tokura$^{1,2,3}$} 
\affiliation{$^1$ Department of Applied Physics and Quantum Phase Electronics Center (QPEC), University of Tokyo, Tokyo 113-8656, Japan \\ $^2$  Cross-Correlated Materials Research Group (CMRG) and Correlated Electron Research Group (CERG), RIKEN Advanced Science Institute, Wako 351-0198, Japan \\  $^3$  Multiferroics Project, ERATO, Japan Science and Technology Agency (JST), Tokyo 113-8656, Japan}

\date{}

\begin{abstract}

Dielectric properties were investigated under various magnitudes and directions of magnetic field ($H$) for a chiral magnetic insulator Cu$_2$OSeO$_3$. We found that the skyrmion crystal induces electric polarization ($P$) along either in-plane or out-of-plane direction of the spin vortices depending on the applied $H$-direction. The observed $H$-dependence of $P$ in ferrimagnetic, helimagnetic, and skyrmion crystal state can be consistently described by the $d$-$p$ hybridization model, highlighting an important role of relativistic spin-orbit interaction in the magnetoelectric coupling in Cu$_2$OSeO$_3$. Our analysis suggests that each skyrmion particle can locally carry electric dipole or quadrupole, which implies that the dynamics of skyrmions are controllable by the external electric field.

\end{abstract}
\maketitle

\begin{figure}
\begin{center}
\includegraphics*[width=8.5cm]{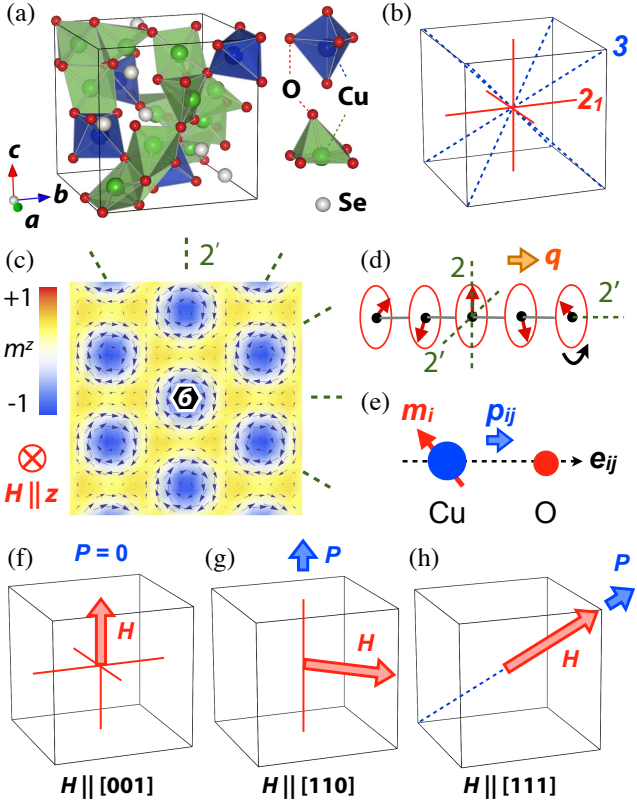}
\caption{(Color Online). (a) Crystal structure of {\COSO} with two distinct Cu$^{2+}$ sites with different oxygen coordination. (b) Symmetry elements compatible with the crystal lattice of {\COSO}. (c) Magnetic skyrmion crystal formed within a plane normal to applied magnetic field ($H$). Background color represents the out-of-plane component of local magnetization vector ($m_z$). (d) Proper screw helical spin texture with a magnetic modulation vector $q$. In (c) and (d), symmetry elements compatible with each spin texture are also indicated; Green dashed lines represent two-fold rotation axes ($2$) or two-fold rotation axes followed by time-reversal ($2'$), and a small black hexagon does a six-fold rotation axis ($6$) along the out-of-plane direction. (e) Schematic illustration of $d$-$p$ hybridization mechanism. (f)-(h) Magnetically-induced electric polarization ($P$) under various directions of $H$ for {\COSO}, predicted by the symmetry analysis (see text).}
\end{center}
\end{figure}

\begin{figure}
\begin{center}
\includegraphics*[width=8.5cm]{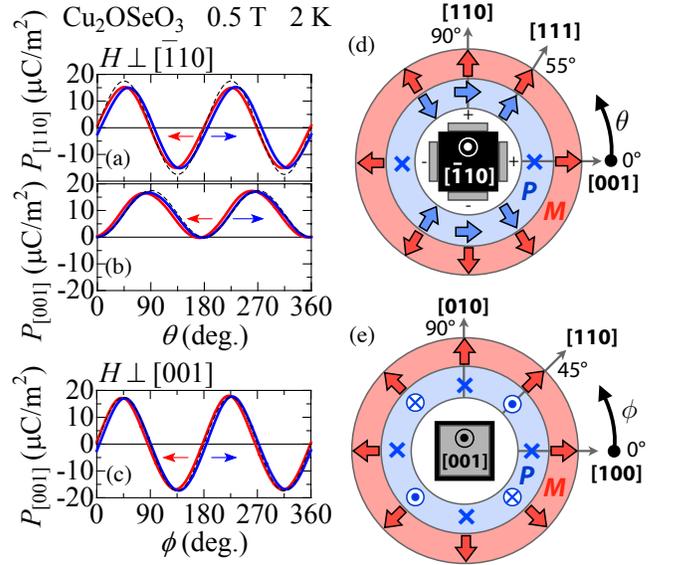}
\caption{(Color Online). (a) [110] and (b) [001] components of $P$ simultaneously measured in $H$ rotating around the $[\bar{1}10]$ axis. (c) [001] component of $P$ under $H$ rotating around the [001] axis. Both measurements were performed at 2 K with $H = 0.5$ T (i.e. collinear ferrimagnetic state). Dashed lines indicate the theoretically expected behaviors based on Eq. (1), and arrows denote the direction of $H$-rotation. In (d) and (e), the experimentally obtained relationships between the directions of $P$ and $M$ in the ferrimagnetic state, as well as the definition of $\theta$ and $\phi$ (the angle between the $H$-direction and the specific crystal axis), are summarized. Here, the directions of $M$ and $P$ are indicated, while $\times$ denotes $P=0$.}
\end{center}
\end{figure}

\begin{figure*}
\begin{center}
\includegraphics*[width=16cm]{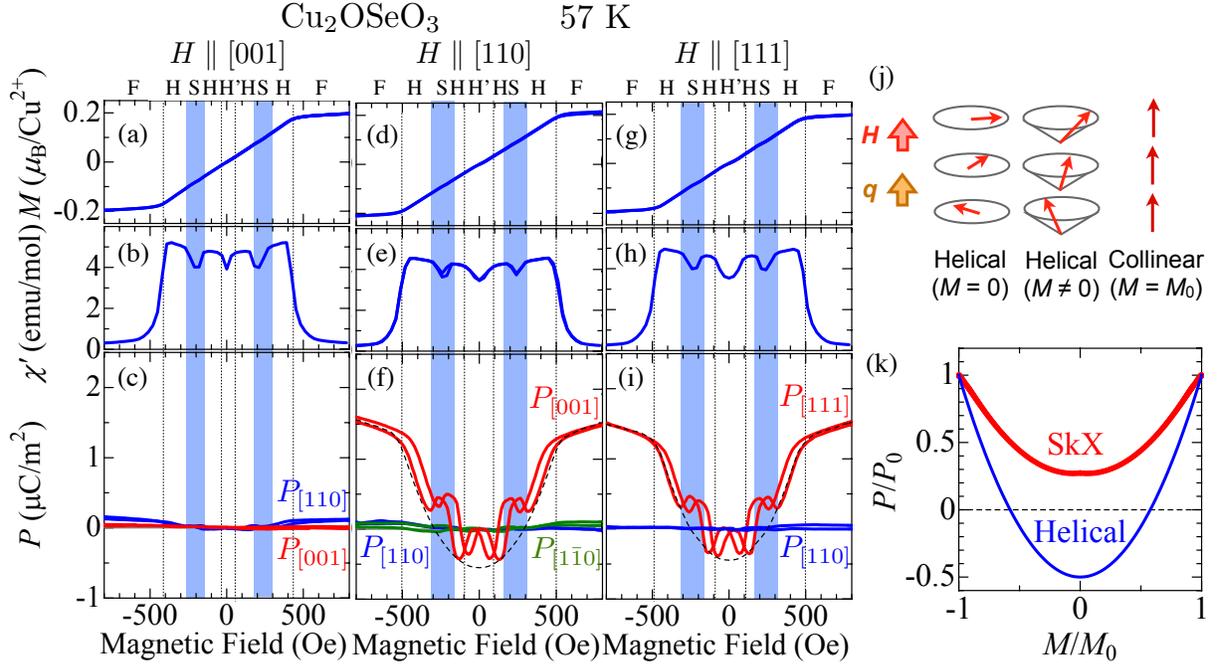}
\caption{(Color Online). Magnetic field dependence of (a) magnetization $M$, (b) ac magnetic susceptibility $\chi '$, and (c) electric polarization $P$ for {\COSO} measured at 57 K with $H \parallel [001]$. The corresponding profiles for $H \parallel [110]$ and $H \parallel [111]$ are also indicated in (d)-(f) and (g)-(i), respectively. Letter symbols F, S, H, and H' stand for ferrimagnetic, SkX, helimagnetic (single $q$-domain) and helimagnetic (multiple $q$-domains) states, respectively. Dashed lines indicate the theoretically expected behavior based on Eq. (1) for the single domain helical spin state. (j) Development of helical spin structure under applied $H$. (k) $P$-$M$ correspondence for the helical and SkX states in {\COSO}, calculated based on the $d$-$p$ hybridization model (see text).}
\end{center}
\end{figure*}

\begin{figure}
\begin{center}
\includegraphics*[width=8.5cm]{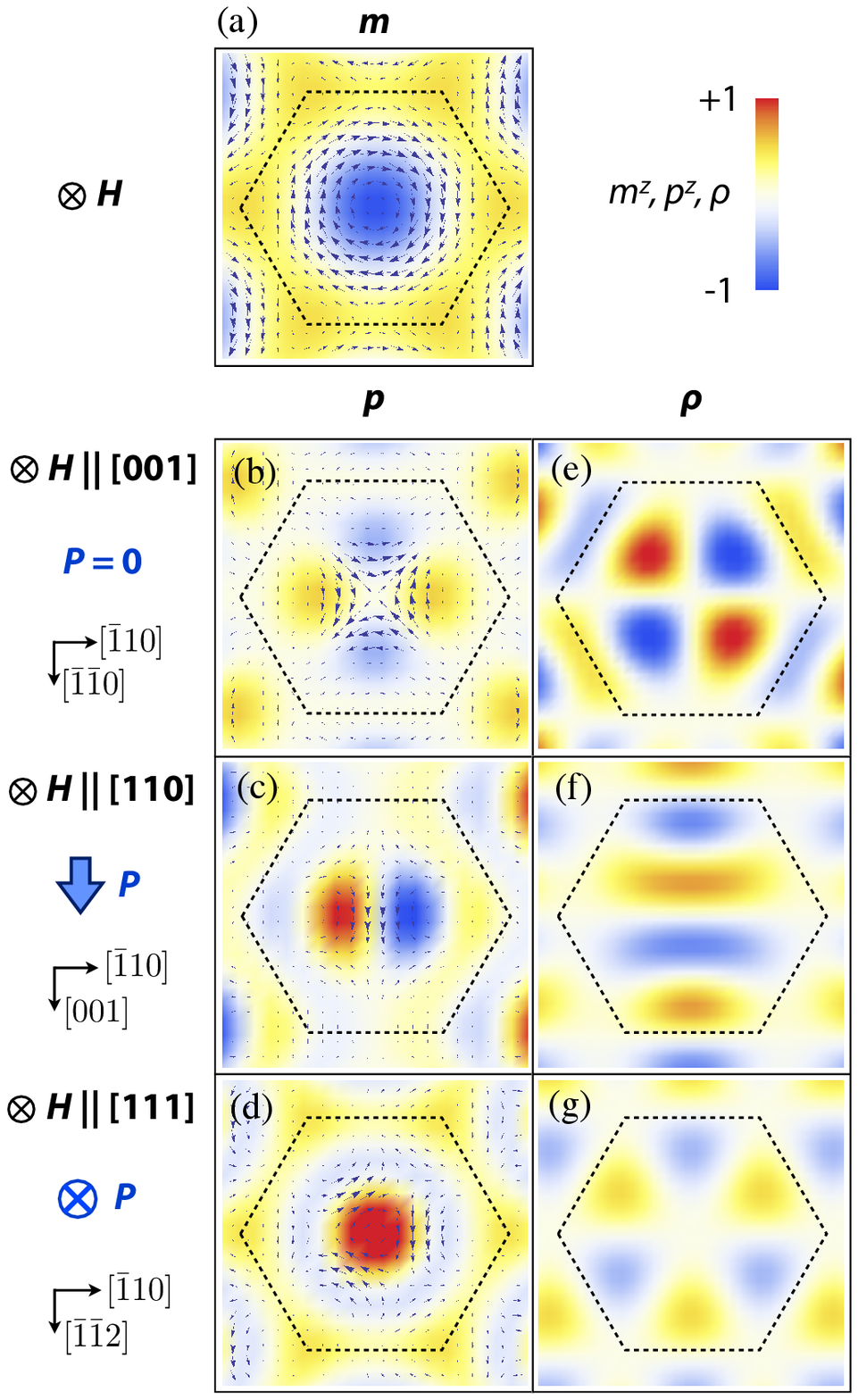}
\caption{(Color Online). Calculated spatial distribution of (a) local magnetization vector $\vec{m}$, (b)-(d) local electric polarization vector $\vec{p}$, and (e)-(g) local electric charge $\rho$ for the skyrmion crystal state (see text). Magnetic field is along the out-of-plane direction, and the results for $H \parallel [001]$ ((b),(e)), $H \parallel [110]$ ((c),(f)), and $H \parallel [111]$ ((d),(g)) are indicated. The background color represents relative value of $m_z$ for (a), $p_z$ for (b)-(d), and $\rho$ for (e)-(g), respectively. Here, $m_z$ and $p_z$ stand for the out-of-plane component of $\vec{m}$ and $\vec{p}$, respectively. The dashed hexagon indicates a magnetic unit cell of the skyrmion crystal, or a single skyrmion particle.}
\end{center}
\end{figure}

Skyrmion is a vortex-like spin swirling object with particle nature\cite{Skyrme, SkXTheory, SkXBogdanov}, which has recently been identified in chiral magnets\cite{NeutronMnSi, TEMFeCoSi}. In metallic system, skyrmions can respond to the electric current through spin transfer torques from conduction electrons\cite{STTMnSi, EmergentMnSi, THEMnSi}; Such electric controllability, as well as its partcile-like nature and nanometric size, highlights skyrmion as a promising building block for next generation of magnetic storage device\cite{Skyrmionics}. In chiral magnets, magnetic interaction acting on two neighboring spins $\vec{S}_i$ and $\vec{S}_j$ consists of two terms; $\vec{S}_i \cdot \vec{S}_j$-like exchange interaction and $\vec{S}_i \times \vec{S}_j$-like Dzyaloshinskii-Moriya (D-M) interaction. Skyrmions are stabilized in the limited window of temperature ($T$) and magnetic field ($H$) over the helimagnetic ground state, and can appear as independent particle or in hexagonal lattice form within a plane normal to $H$ (Fig. 1(c))\cite{NeutronMnSi, TEMFeCoSi}. While the previous observation of skyrmions has been limited to specific metallic alloys with chiral B20 structure such as MnSi\cite{NeutronMnSi}, FeGe\cite{TEMFeGe}, and Fe$_{1-x}$Co$_x$Si\cite{TEMFeCoSi, NeutronFeCoSi}, recently the formation of skyrmion crystal (SkX) has also been identified in an insulator {\COSO} through Lorentz transmission electron microscopy experiments on thin film\cite{Cu2OSeO3_Seki} as well as small angle neutron scattering study on bulk crystal\cite{Cu2OSeO3_SANS_Seki, Cu2OSeO3_SANS_Pfleiderer}.

The crystal structure of {\COSO} belongs to the chiral cubic space group $P2_13$ (Fig. 1(a))\cite{COSO_First, COSO_Structure, COSO_Dielectric}. The magnetic ground state is proper screw, where spins rotate within a plane normal to the magnetic modulation vector $q$ (Fig. 1(d)). In the bulk {\COSO}, the SkX state is stabilized in the narrow $T$- and $H$-region called 'A-phase' just below magnetic ordering temperature $T_c$\cite{Cu2OSeO3_Seki, Cu2OSeO3_SANS_Seki, Cu2OSeO3_SANS_Pfleiderer}. The magnetic modulation period in the helical and SkX state ($\sim 630$ \AA \cite{Cu2OSeO3_SANS_Seki, Cu2OSeO3_SANS_Pfleiderer}) is much longer than the crystallographic lattice constant ($\sim 8.9$ \AA \cite{COSO_Structure, COSO_Dielectric}), and three-up one-down type local ferrimagnetic arrangement between two inequivalent Cu$^{2+}$ sites has been proposed\cite{COSO_Dielectric, COSO_Ferri, COSO_NMR}. Notably, {\COSO} hosts magnetically-induced electric polarization ($P$) in the ferrimagnetic, helimagnetic, and SkX state\cite{Cu2OSeO3_Seki}. The coupling between helical spin texture and ferroelectricity has recently been reported for several compounds like TbMnO$_3$\cite{ME_TbMnO3, ME_Review_Fiebig, ME_Review_Cheong, ME_Review_Seki}, which often enables unique magnetoelectric (ME) response such as a control of magnetism by external electric field ($E$)\cite{ME_PN_LiCu2O2, ME_QVec, ME_MagneticDomain}. Although such a ME control has never been applied to particle-like spin object apart from the current-drive of spin vortex motion in a microdisk magnet, the reported emergence of $P$ in the SkX state implies the possible manipulation of skyrmions by external electric fields in insulators. At this stage, the microscopic origin of ME coupling in {\COSO} has not been identified\cite{COSO_Dielectric, COSO_InfraRed}, and the dielectric properties in the SkX state were investigated only for $H \parallel [111]$\cite{Cu2OSeO3_Seki}. To fully address the possibility of $E$-control of skyrmions in insulators, further study of their ME properties is highly desirable.

In this Letter, we investigated the dielectric properties of bulk {\COSO} single crystal under various magnitudes and directions of $H$. The observed development of $P$ in all magnetic phases can be well reproduced by the recently proposed $d$-$p$ hybridization mechanism\cite{ME_Jia1, ME_Jia2, Ba2MGe2O7_Many}, suggesting an important role of spin-orbit interaction on the ME coupling in {\COSO}. In the crystallized form, skyrmions are found to induce $P$ along the in-plane or out-of-plane direction of spin vortices depending on the $H$-direction with respect to the crystallographic axes. Our analysis shows that each skyrmion particle locally carries electric dipole or quadrupole, which suggests that the dynamics of an individual skyrmion particle is controllable by the external electric field in insulators.

Single crystals of {\COSO} were grown by the chemical vapor transport method\cite{COSO_InfraRed}. They were cut into a cubic shape with faces parallel to (110), (111) or (001), on which silver paste was painted as electrodes. To deduce $P$, we measured the polarization current at constant rates of $H$-sweep (10.6 Oe/sec) or $H$-rotation (1$^\circ$/sec), and integrated it over time. Here, polarization current was measured with an electrometer without applied $E$. Since the transition from paraelectric to ferroelectric state in the process of zero-field-cooling (ZFC) should produce equal population of $P$-domains with no net polarization, we assumes $P = 0$ for the initial $H = 0$ state just after the cooling without $E$ and $H$. Magnetization $M$ and ac magnetic susceptibility $\chi '$ (at 700 Hz) were measured with a SQUID magnetometer.

To investigate the microscopic origin of ME coupling in {\COSO}, we first measured the $H$-direction dependence of $P$ in the collinear (i.e. ferrimagnetic) spin state at 2 K with $H = 0.5$ T. Figures 2(a) and (b) indicate the [110] and [001] component of $P$ ($P_{[110]}$ and $P_{[001]}$) as a function of $H$-direction, simultaneously measured with two sets of electrodes. Here, $H$ rotates around the $[\bar{1}10]$-axis, and $\theta$ is defined as an angle between $H$-direction and the [001] axis (Fig. 2(d)). We also measured the development of $P_{[001]}$ for $H$ rotating around the [001] axis (Fig. 2(c)), where an angle between $H$ and the [100] axis is defined as $\phi$ (Fig. 2(e)). As a function of respective $H$-rotation angle, both $P$-profiles show sinusoidal modulation with the period of $180^\circ$. 

Recently, Jia {\etal} suggested that at least three microscopic mechanisms can be considered as the sources of ME coupling \cite{ME_Jia2, ME_Jia1}. Two of them originate from the correlation between two neighboring magnetic moments $\langle \vec{m}_i \rangle$ and $\langle \vec{m}_j \rangle$ ($\langle \cdots \rangle$ denoting the expected value in the magnetically ordered state), and they provides local electric polarization proportional to $\langle \vec{m}_i \rangle \cdot \langle \vec{m}_j \rangle$ and $\langle \vec{m}_i  \rangle \times \langle \vec{m}_j \rangle$\cite{ME_Katsura}, respectively. However, these two mechanisms predict $P=0$ for the collinear spin state in {\COSO}, due to the cubic symmetry of crystal lattice or to the relationship $ \langle \vec{m}_i  \rangle \times  \langle \vec{m}_j  \rangle = 0$. The third mechanism of ME coupling, namely the $d$-$p$ hybridization model\cite{ME_Jia2, ME_Jia1, Ba2MGe2O7_Many, ME_Arima}, arises from the interaction between a ligand (oxygen) ion and a transition metal (copper) ion with a single magnetic moment $\langle \vec{m}_i \rangle$ (Fig. 1(e)). This model predicts the local electric dipole $\vec{p}_{ij}$ in the form of $\vec{p}_{ij} \propto (\vec{e}_{ij} \cdot \langle \vec{m}_i \rangle)^2 \vec{e}_{ij}$ \cite{Ba2MGe2O7_Many}, where $\vec{e}_{ij}$ is a unit vector along the bond direction\cite{Comment1}. In this scheme, the covalency between metal $d$ and ligand $p$ orbitals is modulated depending on the local magnetization direction via spin-orbit interaction, and thus the local electric dipole is produced along the bond direction\cite{Comment2}. For the given coarse-grained spin texture $\vec{m} (\vec{r})$, the $d$-$p$ hybridization scheme approximately predicts $P$ and $M$ in the form of
\begin{equation}
\vec{P} \propto \frac{1}{\int \textrm{d}\vec{r}}  \int  \sum_{i,j}  (\vec{e}_{ij} \cdot \vec{m} (\vec{r}))^2 \vec{e}_{ij} \textrm{d}\vec{r}
\label{EqP}
\end{equation}
and $\vec{M} \propto \int \vec{m}(\vec{r}) \textrm{d}\vec{r} / \int \textrm{d}\vec{r}$, respectively. Here, summation is taken over 80 Cu-O bonds within a crystallographic unit cell and integral is over a magnetic unit cell. By assuming the collinear spin state with spatially uniform $\vec{m} (\vec{r})$, Eq. (\ref{EqP}) predicts $(P_{[110]}, P_{[001]}) = (2\bar{P}  \sin 2\theta, \bar{P} (1-\cos 2\theta) )$ and $P_{[001]} = 2\bar{P} \sin 2\phi$, where $\bar{P}$ is a common amplitude of $P$. These formulas (dashed lines in Figs. 2(a)-(c)) well reproduce the experimentally observed $H$-direction dependence of $P$, which strongly suggests the validity of $d$-$p$ hybridization mechanism as the origin of ME coupling in {\COSO}. The obtained relationship between the directions of $P$ and $M$ in the collinear spin state is summarized in Figs. 2(d) and 2(e). We can see $P = 0$ for $M \parallel [001]$, while $P \parallel [001]$ for $M \parallel [110]$ and $P \parallel [111]$ for $M \parallel [111]$.

Next, we measured $M$ and $\chi'$ as functions of the magnitude of $H$ at 57 K just below $T_c \sim 58$ K, for $H \parallel [001]$ (Figs. 3(a) and 3(b)), $H \parallel [110]$ (Figs. 3(d) and 3(e)) and $H \parallel [111]$ (Figs. 3(g) and 3(h)). In the ground state ($H=0$), proper screw spin order is realized with mutiple $q$-domains with equivalent $q \parallel \langle 001 \rangle$ directions due to high symmetry of cubic lattice\cite{Cu2OSeO3_Seki, Cu2OSeO3_SANS_Seki, Cu2OSeO3_SANS_Pfleiderer}. Application of $H$ first induces the alignment of $q$ along $H$ direction while keeping the screw-like spin structure, as detected as an enhancement of $\chi '$-value around $H = 100$ Oe. Further increase of $H$ induces continuous transformation of spin texture from proper screw to conical, and finally to colllinear (Fig. 3(j)). It causes almost-linear increase and saturation of $M$-value. In the intermediate field region with $200 < H < 300$ Oe, $\chi '$ shows clear dip anomaly that signals the formation of the SkX state \cite{Cu2OSeO3_Seki, Cu2OSeO3_SANS_Seki, Cu2OSeO3_SANS_Pfleiderer}.

Figures 3 (c), (f) and (i) indicate the corresponding development of $P$ under various directions of $H$. We found that all of ferrimagnetic, helimagnetic, and SkX states induces $P \parallel [111]$ for $H \parallel [111]$, $P \parallel [001]$ for $H \parallel [110]$, and $P = 0$ for $H \parallel [001]$. Such a relationship between $P$ and $H$ can be understood from the viewpoint of symmetry. Whereas the original crystal lattice of {\COSO} belongs to non-polar space group $P2_13$ (Fig. 1(b)), the combination with the ferrimagnetic, proper screw (Fig. 1(d)), or SkX (Fig. 1(c)) spin texture leads to symmetry reduction and may produce an electrically polar state depending on the $H$-direction. In case of $H \parallel [111]$, only a three-fold rotation or screw axis along $H$ remains unbroken and hence $P \parallel H \parallel [111]$ may emerge (Fig. 1(h)) \cite{Cu2OSeO3_Seki}. For $H \parallel [110]$, only a screw (or screw with time-reversal) axis along the [001] direction survives and hence $P \parallel [001]$ is allowed (Fig. 1(g)). With $H \parallel [001]$, the orthogonal arrangement of screw (or screw with time-reversal) axes remains, therefore no $P$ is expected (Fig. 1(f)). Such a situation is realized for all of ferrimagnetic (with $H \parallel M$), proper screw (with $H \parallel q$), and SkX states. The above predictions by the symmetry analysis are fully consistent with the experimental observation.

Based on Eq. (\ref{EqP}), we have further calculated the $P$-$M$ profile for various spin textures in {\COSO}. In the helimagnetic phase, $H$ induces continuous deformation of spin texture from helical to collinear as shown in Fig. 3 (j). The spin texture in this process is described by
\begin{equation}
\vec{m} (\vec{r}) \propto  \vec{e}_z M_{1}  +  [ \vec{e}_{x} \cos (\vec{q} \cdot \vec{r}) + \vec{e}_{y} \sin (\vec{q} \cdot \vec{r}) ],
\label {EqHelTwo}
\end{equation}
where $\vec{e}_{x}$, $\vec{e}_{y}$ and $\vec{e}_{z}$ are unit vectors orthogonal to each other. $\vec{e}_z$ and $\vec{q}$ are parallel to the applied $H$-direction. Likewise, the SkX phase (Fig. 1(c)) is approximately given by\cite{NeutronMnSi}
\begin{equation}
\vec{m} (\vec{r}) \propto \vec{e}_z M_{2}  + \sum ^3_{a = 1} [ \vec{e}_z \cos (\vec{q}_a \cdot \vec{r} + \pi)  + \vec{e}_{a} \sin (\vec{q}_a \cdot \vec{r} + \pi) ].
\label{EqSkyrmionTwo}
\end{equation}
$\vec{q}_a$ denotes one of three magnetic modulation vectors normal to $H$, which forms an angle of 120$^\circ$ with respect to each other. $\vec{e}_{a}$ is a unit vector orthogonal to $\vec{e}_z$ and $\vec{q}_a$, defined so that all $\vec{q}_a \cdot (\vec{e}_z \times \vec{e}_{a}$) may have the same sign. Here, $M_{1}$ and $M_{2}$ scale with the relative magnitude of net magnetization along the $H$-direction, and $M_{1} \rightarrow \infty$ and $M_{2} \rightarrow \infty$ correspond to the collinear spin state; The prefactors for Eq. (\ref{EqHelTwo}) and (\ref{EqSkyrmionTwo}) are determined so as to keep the averaged spin density ($\int |\vec{m}(\vec{r})| \textrm{d}\vec{r} / \int \textrm{d}\vec{r}$) fixed. When we define the $\vec{P}$ and $\vec{M}$ in the collinear spin state as $\vec{P}_0$ and $\vec{M}_0$, Eq. (\ref{EqP}) predicts the relationship $\vec{P} / \vec{P}_0 = \frac{3}{2} (\vec{M}/\vec{M}_0)^2 -  \frac{1}{2}$ for the helimagnetic phase (Eq. (\ref{EqHelTwo})) as shown in Fig. 3 (k). In Figs. 3(f) and 3(i), the experimentally observed $P$-$H$ profiles are fitted with the combination of the calculated $P$-$M$ correspondence (Fig. 3 (k)) and the measured $M$-$H$ profiles (Figs. 3(d) and 3(g)). The fitting curves (dashed lines) well reproduce the observed $P$-$H$ profiles in the helimagnetic phase, including the sign change of $P$. $P$ remains zero in the low-$H$ region with multiple $q$-domains, probably due to the cancelation of $P$ averaged over different domains. Likewise, we also calculated the $P$-$M$ correspondence for the SkX phase with Eq. (\ref{EqSkyrmionTwo}) (Fig. 3(k)). Our calculation predicts $P = 0.41 P_0$ for $M = 0.46 M_0$, which roughly agrees with the experimentally observed values; $P = 0.32 P_0$ for $H \parallel [110]$ (Fig. 3(f)) and $P = 0.33 P_0$ for $H \parallel [111]$ (Fig. 3(i)) for the corresponding $M$-value in the SkX state\cite{Comment3}. The good agreement between the calculated and observed $P$-profiles, including the reproduction of complicated sign change of $P$, confirms that the $d$-$p$ hybridization mechanism adopted here as the microscopic model is responsible for magnetically-induced $P$ in this material.

Since the long magnetic modulation period\cite{Cu2OSeO3_Seki, Cu2OSeO3_SANS_Seki, Cu2OSeO3_SANS_Pfleiderer} in {\COSO} means the almost collinear spin arrangement within a crystallographic unit cell, the experimentally obtained $P$-$M$ correspondence in the collinear spin state (Fig. 2) should also define the relationship between local electric polarization and local magnetization in the SkX state. Figure 4 indicates the spatial distribution of local magnetization $\vec{m}(\vec{r})$ (Fig. 4(a)), as well as local electric polarization $\vec{p} (\vec{r}) \propto \sum_{ij} (\vec{e}_{ij} \cdot \vec{m}(\vec{r}))^2 \vec{e}_{ij}$ (Figs. 4(b)-(d)) and local electric charge $\rho(\vec{r}) \propto \nabla \cdot \vec{p}(\vec{r})$ (Figs. 4(e)-(g)), calculated for the SkX state given by Eq. (\ref{EqSkyrmionTwo}) with various directions of $H$. The obtained $\vec{p} (\vec{r})$ and $\rho(\vec{r})$ profiles suggest that each skyrmion particle locally carries electric quadrupole moment for $H \parallel [001]$ (Figs. 4(b) and 4(e)), or electric dipole moment along the in-plane ([001]) and out-of-plane ([111]) direction of spin vortices for $H \parallel [110]$ (Figs. 4(c) and 4(f)) and $H \parallel [111]$ (Figs. 4(d) and 4(g)), respectively. These pictures are consistent with $P = 0$ for $H \parallel [001]$, $P \parallel [001]$ for $H \parallel [110]$, and $P \parallel [111]$ for $H \parallel [111]$ experimentally obtained by the macroscopic measurement in the SkX state (Figs. 3(c), 3(f), and 3(i)). Such a local coupling between a skyrmion and electric dipole strongly suggests that each skyrmion particle in {\COSO} can be independently driven by the spatial gradient of external electric field. The net charge is always zero within a single skyrmion (Figs. 4(e)-(g)), which implies the non-dissipative nature of skyrmion motion.

In summary, we have investigated the magnetoelectric response of {\COSO} under various magnitudes and directions of magnetic field. The observed development of $P$ in ferrimagnetic, helimagnetic, and skyrmion crystal spin state can be consistently explained by the $d$-$p$ hybridization model, highlighting an important role of relativistic spin-orbit interaction on the magnetoelectric coupling in {\COSO}. In the crystallized form, skyrmions are found to induce electric polarization along either in-plane or out-of-plane direction of the spin vortices depending on the applied $H$-direction. Thus, the skyrmion crystal state in the insulating chiral magnet is endowed with the density-wave nature of both spin and polarization. Our analysis shows that each skyrmion particle can locally carry electric dipole or quadrupole, which strongly suggests the possible manipulation of individual skyrmion particles by external electric field (not current).  Since electric field in insulators causes only negligible joule heat loss compared to the current-driven approach in the metallic system, the presently established "magnetoelectric" skyrmion may contribute to the design of novel spintronic devices with high-energy efficiency.

The authors thank T. Arima, N. Nagaosa, M. Mochizuki, X. Z. Yu, N. Kanazawa, T. Kurumaji, K. Shibata, and M. Rikiso for enlightening discussions and experimental helps. This work was partly supported by FIRST Program by the Japan Society for the Promotion of Science (JSPS).


\begin{thebibliography}{100}
\bibitem{Comment1} Since $\vec{p}_{ij}$ remains unchanged for the reversal of $\vec{m}_i$, the local ferrimagnetic nature of spin texture doesn't affect the calculation of $\vec{p}_{ij}$.
\bibitem{Comment2} In the $d$-$p$ hybridization model, the orbital angular momentum coupled to the spin moment is relevant to the induction of electric polarization, and therefore the approximate treatment of $\langle \vec{m}_i \rangle$ as a classical vector can be justified even for the $S=1/2$ system. The validity of this approach has recently been proved for Ba$_2$CuGe$_2$O$_7$ \cite{Ba2MGe2O7_Many}.
\bibitem{Comment3} To consider the shrinkage of effective magnetic moment near $T_c$, we adopt the $P$- and $M$-values observed at the phase boundary between the ferrimagnetic and helimagnetic phases as $P_0$ and $M_0$.
\end{thebibliography}
\end{document}